\documentclass[aps,twocolumn,prb,preprintnumbers,amsmath,amssymb,superscriptaddress,floats]{revtex4-2}

\usepackage{graphicx}% Include figure files
\usepackage{bm}% bold math
\usepackage{natbib}
\usepackage{float}
\usepackage{multirow}
\usepackage{array}
\usepackage{physics}
\usepackage{siunitx}

\begin{document}

\title{Importance of dynamic lattice effects for crystal field excitations in quantum spin ice candidate Pr$_2$Zr$_2$O$_7$.}

\author{Yuanyuan Xu}
\affiliation{Institute for Quantum Matter and Department of Physics and Astronomy, Johns Hopkins University, Baltimore, Maryland 21218, USA}

\author{Huiyuan Man}
\affiliation{Institute for Quantum Matter and Department of Physics and Astronomy, Johns Hopkins University, Baltimore, Maryland 21218, USA}
\affiliation{Institute for Solid State Physics, University of Tokyo, Kashiwa, Chiba 277-8581, Japan}

\author{Nan Tang}
\affiliation{Institute for Quantum Matter and Department of Physics and Astronomy, Johns Hopkins University, Baltimore, Maryland 21218, USA}
\affiliation{Institute for Solid State Physics, University of Tokyo, Kashiwa, Chiba 277-8581, Japan}

\author{Santu Baidya}
\affiliation{Department of Physics and Astronomy, Rutgers University}

\author{Honbin Zhang}
\affiliation{Department of Physics and Astronomy, Rutgers University}

\author{Satoru Nakatsuji}
\affiliation{Institute for Solid State Physics, University of Tokyo, Kashiwa, Chiba 277-8581, Japan}
\affiliation{CREST, Japan Science and Technology Agency, Kawaguchi, Saitama 332-0012, Japan}
\affiliation{Institute for Quantum Matter and Department of Physics and Astronomy, Johns Hopkins University, Baltimore, MD 21218, USA}
\affiliation{Department of Physics, University of Tokyo, Bunkyo-ku, Tokyo 113-0033, Japan}
\affiliation{Trans-scale Quantum Science Institute, University of Tokyo, Bunkyo-ku, Tokyo 113-0033, Japan}

\author{David Vanderbilt}
\affiliation{Department of Physics and Astronomy, Rutgers University}

\author{Natalia Drichko}
\affiliation{Institute for Quantum Matter and Department of Physics and Astronomy, Johns Hopkins University, Baltimore, Maryland 21218, USA}
\email{Corresponding author. Email:drichko@jhu.edu}

\begin{abstract}
Pr$_2$Zr$_2$O$_7$ is a  pyrochlore  quantum spin-ice candidate. Using Raman scattering spectroscopy we probe crystal electric field excitations of Pr$^{3+}$, and demonstrate the importance of their  interactions with the lattice. We identify a vibronic interaction with a phonon that leads to a splitting of a doublet crystal field excitation at around 55~meV. We also probe a splitting of the non-Kramers ground state doublet of Pr$^{3+}$ by observing a double line of the excitations to the first excited singlet state $E^0_g \rightarrow A_{1g}$. We show that the splitting has a strong temperature dependence, with the doublet structure most prominent  between 50~K and 100~K, and the weight of one of the components strongly decreases on cooling. We suggest a static or dynamic deviation of  Pr$^{3+}$ from the position in the ideal crystal structure can be the origin of the effect, with the deviation strongly decreasing at low temperatures.
\end{abstract}

\date{\today}
\maketitle

\section{Introduction}

Much of condensed matter research is currently focused on a search for an experimental realization of a quantum spin liquid (QSL) state. This is a magnetic state, where spins do not order despite strong interactions, but nevertheless their behavior is determined by strong non-local correlations~\cite{Broholm2020,Savary2016}. It is already understood that this state can be brought about by a presence of a strong geometric frustration or competing interactions. In addition, many candidate systems show some levels of structural disorder. It is  still a question,  if the  disorder prevents  quantum phenomena and mocks them experimentally, or it can it be a factor leading to a QSL state~\cite{Savary2016}. Another important question is how to experimentally distinguish the effects of structural disorder from the effects produced by dynamics of the lattice.  In this work on a quantum spin ice candidate Pr$_2$Zr$_2$O$_7$ \cite{Kimura2013,Wen2017} we show that Raman scattering spectroscopy is able  to separate dynamic effects from the effects of structural disorder. In the case of Pr$_2$Zr$_2$O$_7$ our  study  finds evidence of dynamic lattice effects and  shows their importance for the magnetic state of this material.

Pr$_2$Zr$_2$O$_7$ is a  pyrochlore material, where the crystal structure provides a three dimensional frustrated lattice. Pyrochlores are known to host classical spin ice ~\cite{Gardner2010,Den2000,Melko2001,Bramwell2001} and quantum spin ice, where quantum fluctuations are no longer negligible~\cite{Ross2011,Gingras2014}.  
Magnetic properties of Pr$_2$Zr$_2$O$_7$ are defined by the magnetic moment of Pr$^{3+}$. In a crystal, $^3H_4$ level of $4f$ atomic orbitals of Pr$^{3+}$ which carry $J=4$ magnetic moment is split under the influence of the electric fields related to the local $D_{3d}$ crystal symmetry~\cite{koohpayeh2008,koohpayeh2014} into   $2A_{1g} + A_{2g}+3E_g$ multiplets. This splitting determines magnetic properties of the system. The ground state of the system is the $E_g$ non-Kramers doublet. A non-Kramers doublet  ground state makes the magnetic system sensitive to the lattice degree of freedom, since a small deviation from $D_{3d}$ local symmetry can bring about a splitting of the ground state. This is the basis of suggestions that a local disorder is an important factor in the formation of a quantum spin ice ground state in Pr$_2$Zr$_2$O$_7$~\cite{Wen2017,Martin2017}, as well as a possibility of studies of quantum spin ice by magnetostriction, which were applied to Pr$_2$Zr$_2$O$_7$~\cite{Tang2020,Patri2020}.

The crystal electric field (CEF) description takes into account the lattice degrees of freedom as static, and decoupled from the electronic and magnetic degrees of freedom. This is not always a good approximation. For rare earth materials in particular, the presence of vibronic states, where phonons modulate crystal field levels, is possible, due to the overlapping energy ranges of these excitations. These interactions are not yet widely studied for rare earth pyrochlores, but the example of Tb$_2$Ti$_2$O$_7$, where vibronic coupling is so close to the ground state that it can affect its properties leading to spin liquid state, shows that this dynamic interactions cannot be neglected ~\cite{Constable2017,Zhang2020,Rau2019}.

Here we present our new findings on the crystal field levels of  Pr$^{3+}$ in  Pr$_2$Zr$_2$O$_7$ and their coupling to the lattice. High spectral resolution and symmetry selectivity of Raman scattering spectroscopy allows a new look at the importance of interactions with the lattice, going beyond simple crystal field splitting and the static approximation. We observe a splitting of the doubly-degenerate crystal field levels with the nature of the splitting being different for the $E_g$ levels of different energies. We demonstrate that the $E_g$ level at 55 meV shows a splitting of 2.3 meV due to vibronic interactions missed in previous studies~\cite{Kimura2013,Bonville2016,Martin2017}.  We also probe the $E_g$ ground-state splitting  and its evolution with temperature through the analysis of transitions to the lowest excited $A_{1g}$ state. Our results suggest that the splitting is present prominently at temperatures around 100~K, and decreases on cooling. We discuss possible static and dynamic origins of this effect.

\section{Results}
\begin{figure}[!htb]
	\includegraphics[width=\linewidth]{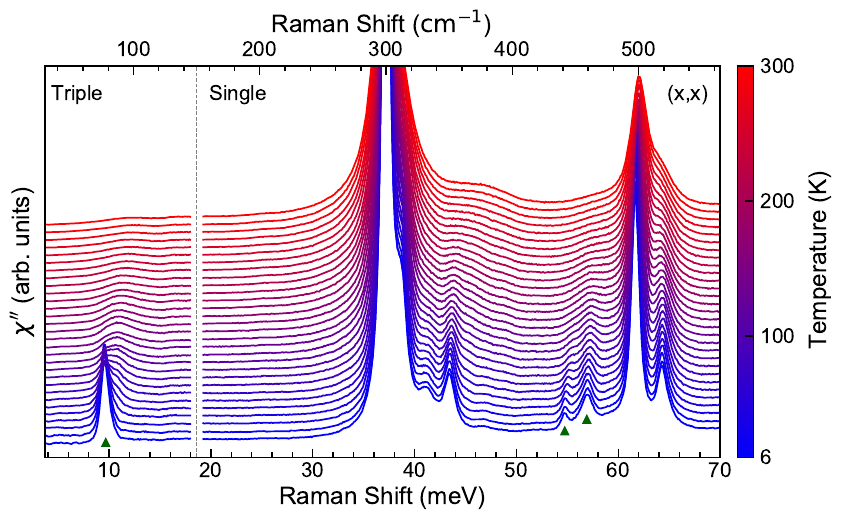}
	\caption{Temperature dependence of the Raman scattering spectra of Pr$_2$Zr$_2$O$_7$ in the temperature range from 6 to 300~K in parallel polarization configuration $(x,x)$ in the spectral region of the CEF. The spectra are shifted along y-axis for clarity. CEF excitations are marked by green triangles. The full measured spectral range up to 125 meV can be found in SI.}
	\label{Fig1}
\end{figure}

The temperature dependence of the Raman spectra of Pr$_2$Zr$_2$O$_7$ in the spectral range from 3.7 to 70~meV (30 to 565~cm$^{-1}$ at temperatures between 6 and 300~K is presented in Fig.~\ref{Fig1}. Raman spectra in the range up to  125 meV can be found in SI. In the Raman spectra of Pr$_2$Zr$_2$O$_7$   we observe two types of excitations: (i) Raman-active phonons; (ii) CEF excitations of Pr$^{3+}$. Phonons were identified by  polarization-resolved Raman measurements on the (100) surface~\cite{Xu2020phonons} and  a comparison to the DFT phonons calculations, while CEF excitations were assigned based on  neutron scattering results~\cite{Kimura2013,Princep2013,Bonville2016}.

In this work  we  focus our attention on the CEF excitations in the spectra of Pr$_2$Zr$_2$O$_7$. Raman scattering can detect spectral lines of CEF excitations with much higher energy resolution (0.125 meV for our experiments) than that of the neutron scattering measurements, which can typically go down only to 1 meV in the high resolution measurements for low signals~\cite{Gaudet2018}. CEF excitations show a much stronger temperature dependence  of the line width than  phonons~\cite{Sanjurjo1994} (see Fig.~\ref{Fig1}), which allows to observe most of them only at low temperatures.

\begin{table}[H]
	\caption{Frequencies and widths  of Pr$^{3+}$ CEF levels obtained from the Raman scattering spectra at $T=14$~K. For the $A_{1g}$ excitation at 9.5 meV, the second component has below 10\% of the spectral weight and is not included into the table. For the CEF excitations around 55 meV, $v_1$ and $v2$ are magnetoelastically induced vibronic states which possess $A_{1g}$ and $E_g$ symmetry respectively.  \label{table:CEF}}
	\centering
	\begin{ruledtabular}
	\begin{tabular}{
		>{\centering\arraybackslash}l
		>{\centering\arraybackslash}m{0.38\linewidth}
		>{\centering\arraybackslash}m{0.38\linewidth}}
	Level & Frequency (meV) & Line width (meV) \\
	\hline
	 $A_{2g}$ & 109.0 & 1.5  \\
	 $E_g$ & 94.4 & 2.8   \\
	 $A_{1g}$ & 82.1 & 1.5 \\
	 $v_2~(E_g)$ & 57.1 & 1.2 \\
	 $v_1~(A_{1g})$ &  54.8 & 0.6 \\
	 $A_{1g}$ & 9.5 & 1.0  \\	
	\end{tabular}
	\end{ruledtabular}
\end{table}

Energies of Pr$^{3+}$ CEF excitations observed in the Raman spectra of Pr$_2$Zr$_2$O$_7$ (see Tab~\ref{table:CEF}) correspond well to that observed by neutron scattering~\cite{Kimura2013,Bonville2016}. However, the line shapes measured with the higher spectral resolution provided a lot of new information. We present spectra of CEF at 14~K with phonons subtracted in Fig.~\ref{Fig2}(a). The main result is the Raman  observation of splitting of the lower energy  doublet levels, and the evidence for the different physical origins of  the splitting of different excitations.

For higher energy CEF states, we observe a  difference in line width between singlet and doublet excitations. At 14~K the spectral line corresponding to the excitation of the doublet $E_g$ at 94 meV (762 cm$^{-1}$) shows a width of 2.8 meV, which  is about two times larger than the width of the spectral lines of singlet $A_{1g}$ and $A_{2g}$ excitations (1.5~meV), see Table~\ref{table:CEF}, Fig~\ref{Fig2}(a).

The  line of a doublet $E_g$  at about 55 meV (460~cm$^{-1}$) is split into two components ($v_1$ and $v_2$) separated by 2.3 meV.   The $v_1$ and $v_2$ components of the excitation show different symmetries, with the low-frequency component $v_1$ following the properties of $xx+yy$ basis functions ($A_{1g}$ scattering channel), and the higher frequency one ($v_2$) following $x^2-y^2$ basis functions ($E_{g}$ scattering channel), see Fig.~\ref{Fig2}a.

On increasing the temperature, the excitation lines broaden (for details see SI, Fig.~3). All the lines of the CEF excitations harden by about 1~meV on the increase of temperature.

At temperatures below about 20~K, the line of the CEF excitation to the lowest excited  singlet $A_{1g}$ level at 9.5~meV (Fig.~\ref{Fig3}) shows an asymmetric shape. It can be well described by two symmetric Gaussian-Lorentzian line shapes with the higher-frequency component of approximately 10 \% of the total spectral weight of the excitation line. On the increase of the temperature, the spectral weight of the high-energy component increases, and the line develops into a well-defined doublet line with overlapping  components at about 9.5~meV and 10.4~meV. The energy difference between the two components of the doublet increases, and lines broaden on temperature increase  (see Fig.~\ref{Fig3}(b)), until  the components cannot be distinguished  above 110~K.  The doublet is identified most distinctly in the temperature range between 50~K and 100~K.  The components of the doublet  do not show any polarization dependence. The slight asymmetry is also present in the line shapes of the higher-frequencies $A_{1g}$ CEF excitations, however, at all temperatures  they are much broader than the energy difference between the two components of the excitation at 9.5 meV.

 \section{Discussion}

According to the average crystal structure~\cite{koohpayeh2014}, the $4f$ level of Pr$^{3+}$ in Pr$_2$Zr$_2$O$_7$ is split by the crystal field of $D_{3d}$ symmetry into  $2A_{1g} + A_{2g} +3E_g$ multiplets. In the Raman scattering spectra we observe excitations from the ground state doublet to all higher-energy components of this multiplet~\cite{Kimura2013}.  However, neither the symmetry-dependent splitting of the $E_g$ level at 55 meV, nor the splitting of the $A_{1g}$ level at 9.5~meV, can be understood within simple picture of the crystal field splitting according to $D_{3d}$ symmetry.

\subsection{Vibronic state for $E_g$ excitation at 55 meV.}
First, we discuss the origin of the splitting of the $E_g$ level at 55~meV. The splitting between the resulting lines (2.3 meV) is larger that the splitting of the ground state doublet (1~meV), the latter estimated according to our data and to previously published neutron scattering data ~\cite{Wen2017,Martin2017}.

The well-defined symmetry of the components shows that the splitting  cannot be understood in terms of structural disorder leading to a relief of the double degeneracy of the level, contrary to the previous interpretation~\cite{Martin2017}.
Such a symmetry-defined splitting  can occur on mixing with another excitation of $E$ symmetry, with $E \otimes E = E + A_1 + A_2$, where the resulting $A_1$ and $E$ excitations will be observed  in $(x,y)$  and $(x,x)$  scattering channels, as detected in our experiment. Thus the candidate excitation should have $E$ symmetry and energy close to the 55 meV of the crystal field level.  In Pr$_2$Zr$_2$O$_7$, similar to other rare-earth based crystals, a good candidate for such an excitation is a phonon.

Theory describing this vibronic process was first developed by P.~Thalmeier et.~al~\cite{Thalmeier1982}.
Typically, this mixing occurs when a CEF state and a phonon have the same symmetry and are close in frequency. The system can be described by the following Hamiltonian~\cite{Thalmeier1982}
\begin{equation}
	\mathcal{H} = \mathcal{H}_0 + \mathcal{H}_{\rm{int}},
	\label{eq:1}
\end{equation}
with the non-interacting part:
\begin{equation}
	\mathcal{H}_0 = \sum_{\alpha n}\epsilon_\alpha \ket{\Gamma_\alpha^n}\bra{\Gamma_\alpha^n} + \hbar\omega_0\sum_\mu (a_\mu^\dagger a_\mu + \frac{1}{2})
	\label{eq:2}
\end{equation}
and the interacting part
\begin{equation}
	\mathcal{H}_{\rm{int}} = -g_0\sum_\mu U_\mu O_\mu.
	\label{eq:3}
\end{equation}
The non-interacting part of the Hamiltonian (Eqn.~\ref{eq:2}) is composed of a CEF level $\ket{\Gamma_\alpha}$ of a rare earth ion with degeneracy index $n$ and a coupled phonon with energy $\hbar\omega_0$. In the interacting part (Eqn.~\ref{eq:3}), $U_\mu = a_\mu + a_\mu^\dagger$ is the phonon displacement operator and $O_\mu$ is the quadrupolar operator transforming like the symmetry of the phonon~\cite{Lovesey2000, Ruminy2017}. The magnetoelastic coupling constant is given by $g_0$. Importantly, this process is not restricted to any particular part of the Brillouin zone.

\begin{figure}[!htb]
	\includegraphics[width=\linewidth]{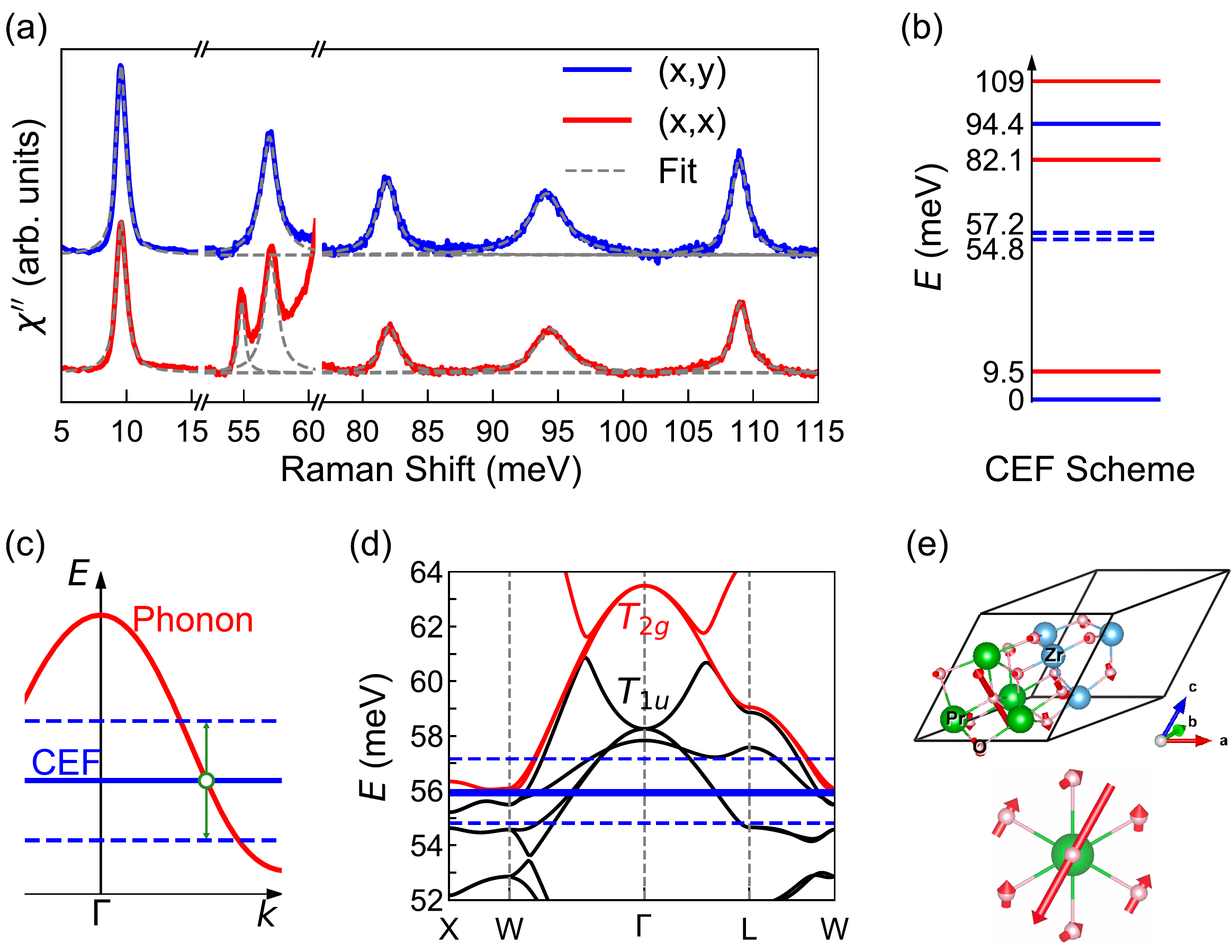}
	\caption{(a) Raman spectra of CEF excitation of Pr$_2$Zr$_2$O$_7$ at 14~K. (b) A scheme of the CEF levels of Pr$^{3+}$ in Pr$_2$Zr$_2$O$_7$. (c) Diagram of the vibronic coupling between the phonons and CEF states. (d) Pr$_2$Zr$_2$O$_7$ phonon dispersion obtained by DFT calculations. The energy range close to the vibronic features is shown. Dashed blue lines mark the experimentally observed CEF vibronic excitations. The vibronic coupling is expected to occur at the intersection point of the nonsplit CEF doublet (solid blue line) and the $T_{2g}$ phonon branch (red curve). (e) Atomic displacements of the $T_{2g}$ phonon mode. Upper panel: view of the unit cell. Lower panel: PrO$_8$ octahedron site viewed from the $[111]$ direction.}
	\label{Fig2}
\end{figure}

In the case of Pr$_2$Zr$_2$O$_7$, we can identify the  phonon which produces the vibronic state. The calculated phonon dispersion in the relevant energy range is presented in Fig~\ref{Fig2}(d).  At the $\Gamma$-point of the BZ, the calculations are in good agreement with the experimentally determined frequencies of the Raman-active phonons~\cite{Xu2020phonons}, and both show an absence of a doubly-degenerate phonon in the region of 55~meV. However, mixing is possible at the other parts of the BZ, which also puts less restrictions on the phonon symmetry. A simple assumption is that the energy of an unperturbed $E_g$ crystal field excitation would be found between the two split components at 56 meV, as marked with  a dashed line in Fig.~\ref{Fig2}. There are two phonon candidates that are very close in energy to this CEF excitation in the X to W part of the BZ. The most probable phonon candidate is a phonon which is observed at 64.6~meV (63.5~meV is the calculated frequency) at the $\Gamma$ point ($T_{2g}$). The calculated dispersion of this phonon is plotted as a red line in Fig.~\ref{Fig2}(d). The eigenvector involves the movement of O1 oxygens, which modulate the Pr$^{3+}$ oxygen environment (see Fig.~\ref{Fig2}(e)).  We observe this vibronic effect in the Raman spectra at $\Gamma$-point, because the CEF excitations do not show dispersion, and the splitting which results from  interactions in a certain part of BZ, leads to the splitting of the CEF levels observed over the whole BZ~\cite{Gaudet2018}.

Our high resolution measurements of CEF excitations allow us to refine the crystal field parameters~\cite{Bonville2016}  and obtain values of magneto-elastic coupling constants, as shown in detail  in the SI.

Rare earth atoms show CEF excitations in the energy range of the lattice phonons in many materials, and the vibronic effect involving the CEF excitations can be  relatively common~\cite{Thalmeier1982,Thalmeier1984,Gaudet2018}, though not broadly studied.  Among pyrochlore rare earth based compounds,  vibronic states  are found, for example, in Ho$_2$Ti$_2$O$_7$~\cite{Gaudet2018} and Tb$_2$Ti$_2$O$_7$~\cite{Constable2018}. The latter is an especially interesting case, because a mixing of the very-low lying first excited state with a phonon might be an origin of a spin liquid state in this material~\cite{Gingras2014}.

\subsection{Probing the splitting of the ground state doublet.}

\begin{figure}[!htb]
	\includegraphics[width=\linewidth]{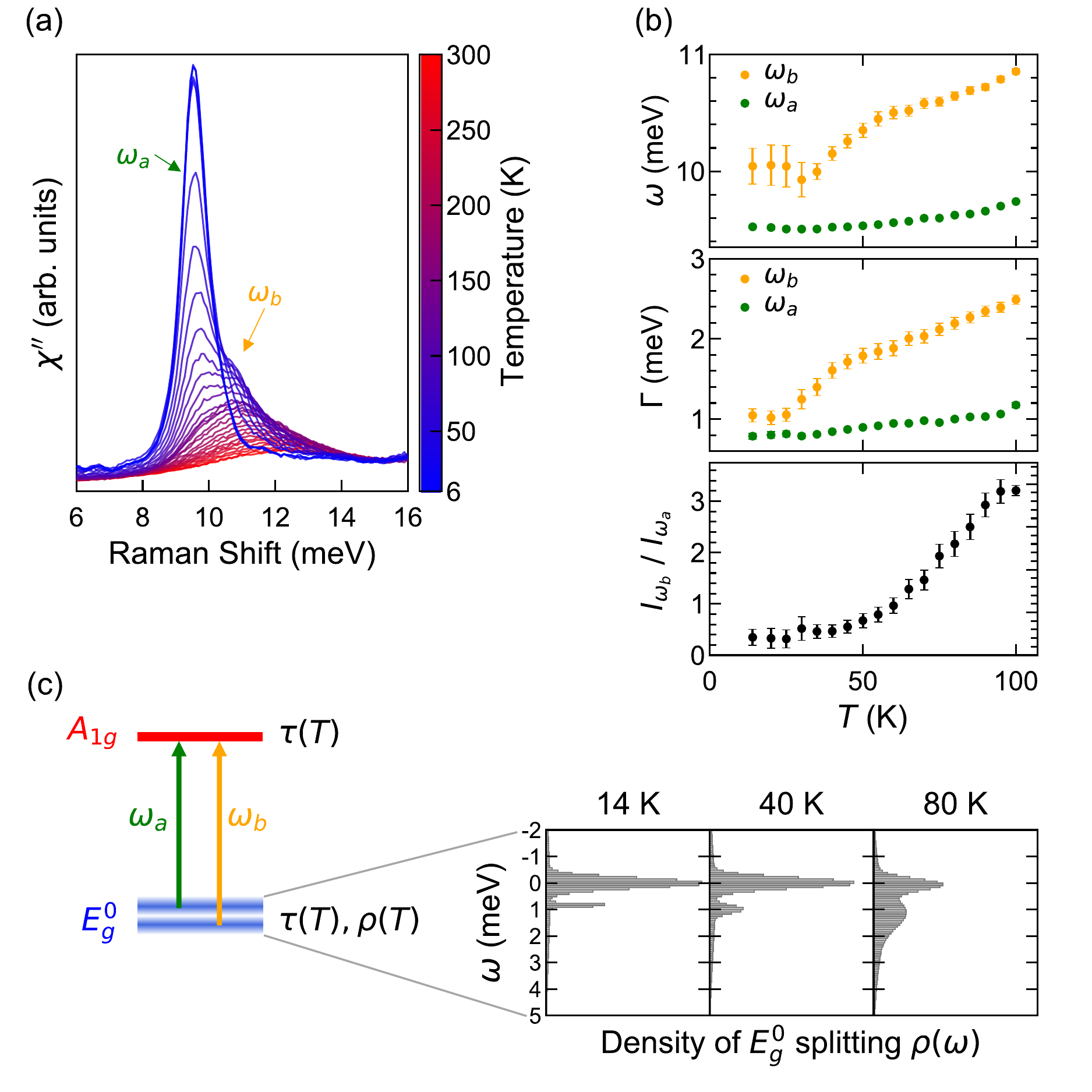}
	\caption{(a) Temperature dependence of the Raman scattering spectra in the energy range of the lowest lying CEF excitation $E_g^0 \rightarrow A_{1g}$. (b) Upper panel: positions of the two components of the transition ($\omega_a$ and $\omega_b$); middle panel: line widths of the two components of the excitations ($\omega_a$ and $\omega_b$);  lower panel: the ratio of the spectral weight of $\omega_b$ to $\omega_a$. (c) The scheme of the excitation $E_g^0 \rightarrow A_{1g}$ illustrates that the shape of the electronic density distribution $\rho$ of the $E^0_g$ level, such as splitting, can define the shape of the observed spectral excitation line. Histograms demonstrate the density of the ground state splitting $\rho(\omega)$ obtained by the deconvolution of the Raman scattring spectra at 14, 40, and 80 K.}
	\label{Fig3}
\end{figure}

The splitting of the $A_{1g}$ CEF excitation at 9.5~meV has a very different character. The doublet line of this excitation does not show any polarization dependence.  This symmetry consideration, together with an absence of the $A_{1g}$ $\Gamma$- point phonons close to 9.5 meV,  allows us to dismiss  a vibronic state interpretation~\footnote{For  $A_{1g}$ singlet level only vibronic interaction with a $\Gamma$-point phonon would produce a doublet}.

A splitting of the spectral line corresponding to the  $E_g^0 \rightarrow A_{1g}$ excitation can reflect the doublet structure of the   $E_g^0$ ground state level, as is schematically shown in Fig.~\ref{Fig3}.  It can be understood easily by considering the relevant  Raman intensity $\chi''(\omega)$ at each frequency $\omega_i$:
	\begin{equation}
		\chi''(\omega_i, T) = B_{nk}\int_{-\infty}^{\infty}\rho(\omega', T)L(\omega_i - \omega', T)d\omega'.
	\end{equation}
	Here the natural width of the $A_{1g}$ singlet level  is a Lorenzian function $L(\omega,T)$ determined by the lifetime of the level $\tau(T)$~\cite{Sanjurjo1994}, and $\rho(\omega,T)$ is the density of the $E^0_g$ level. $B_{nk}$ a probability of $E_g \rightarrow A_{1g}$ transition.

We can use a deconvolution procedure for experimental Raman intensity $\chi(\omega)$ to obtain  $\rho(\omega,T)$. The strong temperature dependence of the shape of the Raman excitation at around 9.5~meV   reflects the change of $\rho(\omega,T)$ with temperature. We show $\rho(\omega,T)$ for a number of temperatures in Fig.~\ref{Fig3}(c). While the  doublet structure is pronounced at temperatures between 100 and 40~K, the  relative spectral weight of the high-frequency component decreases on cooling, reaching only 10\% of the spectral weight of the lower-frequency component below 30 K. This temperature dependence  is reversed to that expected due to the thermal population of the levels. $\rho(\omega_i,T)$ is approaching that of a single non-split $E_g^0$ level as the temperature is reduced. The non-smooth temperature dependence of the parameters of the $E^0_g$ components at around 30 K reflects the decrease of the low-energy weak component down to below 10\% of the total weight. Interestingly, in this temperature range, a change of slope in magnetic susceptibility is observed (see SI) \cite{Kimura2013, Bonville2016}.  At temperatures below about 30 K susceptibility is well-described by a non-split ground state CEF level only \cite{Bonville2016}.

An observation of the $E^{0}_g$ doublet in Raman scattering was possible due the high spectral resolution, and presents a more complicated picture than a single band Gaussian distribution with width of 1~mW obtained from neutron scattering data~\cite{Martin2017,Wen2017}. The latter  was suggested to be a result of structural disorder~\cite{Wen2017}, and in particulary random strain~\cite{Martin2017}.   The presence of the two components of $E^{0}_g$ separated by about 1~mW  can explain a feature in the heat capacity of Pr$_2$Zr$_2$O$_7$ observed at about 10~K ~\cite{Kimura2013}.

While random disorder cannot be the origin of the well-defined splitting, the split ground state doublet can be a result of  a deviation of Pr$^{3+}$ environment from  $D_{3d}$. Such a deviation may be produced, for example, by  a shift of the Pr$^{3+}$ atom from the position in the average structure, and was suggested by B.~Trump et.~al~\cite{Trump2018}. The temperature dependence of the spectra corresponds to a decrease of a deviation of  Pr$^{3+}$ environment from  $D_{3d}$ on cooling, which can occur due to changes of structure on thermal contraction of the crystals.

Alternatively, the splitting can originate from a dynamic process. Such a process could be a phonon-like  ``jump'' of Pr atom between a central position and an off-center potential minimum. Such a picture can explain the temperature dependence of both  intensities and width of the $E_g^0$ components (Fig~\ref{Fig3}). On cooling, this dynamic process slows down or the phonon excitation gets de-populated, leading to the redistribution of spectral weight and narrowing of the levels that belong  to the ground state $E_g^0$ doublet. If such a phonon mode exists, it would be a dipole-active excitation, observed in GHz regime.

\section{Conclusions}

In this work we perform a high-resolution symmetry-resolved Raman scattering study of  crystal field levels of Pr$^{3+}$  in Pr$_2$Zr$_2$O$_7$, and show that dynamic interactions with the lattice are the dominant reason for the splitting of the doublet crystal fields levels. We show that a 2.3~meV splitting of an $E_g^0$ crystal field level at 55~meV originates from a vibronic interaction with a phonon.

We detect a splitting of the ground state doublet by analyzing a transition to the first excited state  $E_g^0 \rightarrow A_{1g}$. The splitting has a strong temperature dependence. We suggest possible interpretations in terms of static or dynamic shift of Pr$^{3+}$ from the $D_{3d}$.

\section{Acknowledgements}

The authors are thankful to J.~Gauget, C.~Broholm, S.~Bhattacharjee, T.~McQueen, J-J.~Wen, and  O.~Tchernyshev for useful discussions. This work was supported as part of the Institute for Quantum Matter, an Energy Frontier Research Center funded by the U.S. Department of Energy, Office of Science, Basic Energy Sciences under Award No.~DE-SC0019331. This work in Japan is partially supported by CREST (Grant Number: JPMJCR18T3 and JPMJCR15Q5), by New Energy and Industrial Technology Development Organization (NEDO), by Grants-in-Aids for Scientific Research on Innovative Areas (Grant Number: 15H05882 and 15H05883) from the Ministry of Education, Culture, Sports, Science, and Technology of Japan, and by Grants-in-Aid for Scientific Research (Grant Number: 19H00650) and Program for Advancing Strategic International Networks to Accelerate the Circulation of Talented Researchers (Grant Number: R2604) from the Japanese Society for the Promotion of Science (JSPS).

\bibliography{PZO}

\providecommand{\noopsort}[1]{}\providecommand{\singleletter}[1]{#1}%
\begin{thebibliography}{30}%
\makeatletter
\providecommand \@ifxundefined [1]{%
 \@ifx{#1\undefined}
}%
\providecommand \@ifnum [1]{%
 \ifnum #1\expandafter \@firstoftwo
 \else \expandafter \@secondoftwo
 \fi
}%
\providecommand \@ifx [1]{%
 \ifx #1\expandafter \@firstoftwo
 \else \expandafter \@secondoftwo
 \fi
}%
\providecommand \natexlab [1]{#1}%
\providecommand \enquote  [1]{``#1''}%
\providecommand \bibnamefont  [1]{#1}%
\providecommand \bibfnamefont [1]{#1}%
\providecommand \citenamefont [1]{#1}%
\providecommand \href@noop [0]{\@secondoftwo}%
\providecommand \href [0]{\begingroup \@sanitize@url \@href}%
\providecommand \@href[1]{\@@startlink{#1}\@@href}%
\providecommand \@@href[1]{\endgroup#1\@@endlink}%
\providecommand \@sanitize@url [0]{\catcode `\\12\catcode `\$12\catcode
  `\&12\catcode `\#12\catcode `\^12\catcode `\_12\catcode `\%12\relax}%
\providecommand \@@startlink[1]{}%
\providecommand \@@endlink[0]{}%
\providecommand \url  [0]{\begingroup\@sanitize@url \@url }%
\providecommand \@url [1]{\endgroup\@href {#1}{\urlprefix }}%
\providecommand \urlprefix  [0]{URL }%
\providecommand \Eprint [0]{\href }%
\providecommand \doibase [0]{https://doi.org/}%
\providecommand \selectlanguage [0]{\@gobble}%
\providecommand \bibinfo  [0]{\@secondoftwo}%
\providecommand \bibfield  [0]{\@secondoftwo}%
\providecommand \translation [1]{[#1]}%
\providecommand \BibitemOpen [0]{}%
\providecommand \bibitemStop [0]{}%
\providecommand \bibitemNoStop [0]{.\EOS\space}%
\providecommand \EOS [0]{\spacefactor3000\relax}%
\providecommand \BibitemShut  [1]{\csname bibitem#1\endcsname}%
\let\auto@bib@innerbib\@empty
%</preamble>
\bibitem [{\citenamefont {Broholm}\ \emph {et~al.}(2020)\citenamefont
  {Broholm}, \citenamefont {Cava}, \citenamefont {Kivelson}, \citenamefont
  {Nocera}, \citenamefont {Norman},\ and\ \citenamefont
  {Senthil}}]{Broholm2020}%
  \BibitemOpen
  \bibfield  {author} {\bibinfo {author} {\bibfnamefont {C.}~\bibnamefont
  {Broholm}}, \bibinfo {author} {\bibfnamefont {R.~J.}\ \bibnamefont {Cava}},
  \bibinfo {author} {\bibfnamefont {S.~A.}\ \bibnamefont {Kivelson}}, \bibinfo
  {author} {\bibfnamefont {D.~G.}\ \bibnamefont {Nocera}}, \bibinfo {author}
  {\bibfnamefont {M.~R.}\ \bibnamefont {Norman}},\ and\ \bibinfo {author}
  {\bibfnamefont {T.}~\bibnamefont {Senthil}},\ }\bibfield  {journal} {\bibinfo
   {journal} {Science}\ }\textbf {\bibinfo {volume} {367}},\ \href
  {https://doi.org/10.1126/science.aay0668} {10.1126/science.aay0668} (\bibinfo
  {year} {2020})\BibitemShut {NoStop}%
\bibitem [{\citenamefont {Savary}\ and\ \citenamefont
  {Balents}(2016)}]{Savary2016}%
  \BibitemOpen
  \bibfield  {author} {\bibinfo {author} {\bibfnamefont {L.}~\bibnamefont
  {Savary}}\ and\ \bibinfo {author} {\bibfnamefont {L.}~\bibnamefont
  {Balents}},\ }\href {https://doi.org/10.1088/0034-4885/80/1/016502}
  {\bibfield  {journal} {\bibinfo  {journal} {Reports on Progress in Physics}\
  }\textbf {\bibinfo {volume} {80}},\ \bibinfo {pages} {016502} (\bibinfo
  {year} {2016})}\BibitemShut {NoStop}%
\bibitem [{\citenamefont {Kimura}\ \emph {et~al.}(2013)\citenamefont {Kimura},
  \citenamefont {Nakatsuji}, \citenamefont {Wen}, \citenamefont {Broholm},
  \citenamefont {Stone}, \citenamefont {Nishibori},\ and\ \citenamefont
  {Sawa}}]{Kimura2013}%
  \BibitemOpen
  \bibfield  {author} {\bibinfo {author} {\bibfnamefont {K.}~\bibnamefont
  {Kimura}}, \bibinfo {author} {\bibfnamefont {S.}~\bibnamefont {Nakatsuji}},
  \bibinfo {author} {\bibfnamefont {J.}~\bibnamefont {Wen}}, \bibinfo {author}
  {\bibfnamefont {C.}~\bibnamefont {Broholm}}, \bibinfo {author} {\bibfnamefont
  {M.}~\bibnamefont {Stone}}, \bibinfo {author} {\bibfnamefont
  {E.}~\bibnamefont {Nishibori}},\ and\ \bibinfo {author} {\bibfnamefont
  {H.}~\bibnamefont {Sawa}},\ }\href {https://doi.org/10.1038/ncomms2914}
  {\bibfield  {journal} {\bibinfo  {journal} {Nature communications}\ }\textbf
  {\bibinfo {volume} {4}},\ \bibinfo {pages} {1934} (\bibinfo {year}
  {2013})}\BibitemShut {NoStop}%
\bibitem [{\citenamefont {Wen}\ \emph {et~al.}(2017)\citenamefont {Wen},
  \citenamefont {Koohpayeh}, \citenamefont {Ross}, \citenamefont {Trump},
  \citenamefont {McQueen}, \citenamefont {Kimura}, \citenamefont {Nakatsuji},
  \citenamefont {Qiu}, \citenamefont {Pajerowski}, \citenamefont {Copley},\
  and\ \citenamefont {Broholm}}]{Wen2017}%
  \BibitemOpen
  \bibfield  {author} {\bibinfo {author} {\bibfnamefont {J.-J.}\ \bibnamefont
  {Wen}}, \bibinfo {author} {\bibfnamefont {S.~M.}\ \bibnamefont {Koohpayeh}},
  \bibinfo {author} {\bibfnamefont {K.~A.}\ \bibnamefont {Ross}}, \bibinfo
  {author} {\bibfnamefont {B.~A.}\ \bibnamefont {Trump}}, \bibinfo {author}
  {\bibfnamefont {T.~M.}\ \bibnamefont {McQueen}}, \bibinfo {author}
  {\bibfnamefont {K.}~\bibnamefont {Kimura}}, \bibinfo {author} {\bibfnamefont
  {S.}~\bibnamefont {Nakatsuji}}, \bibinfo {author} {\bibfnamefont
  {Y.}~\bibnamefont {Qiu}}, \bibinfo {author} {\bibfnamefont {D.~M.}\
  \bibnamefont {Pajerowski}}, \bibinfo {author} {\bibfnamefont {J.~R.~D.}\
  \bibnamefont {Copley}},\ and\ \bibinfo {author} {\bibfnamefont {C.~L.}\
  \bibnamefont {Broholm}},\ }\href
  {https://doi.org/10.1103/PhysRevLett.118.107206} {\bibfield  {journal}
  {\bibinfo  {journal} {Phys. Rev. Lett.}\ }\textbf {\bibinfo {volume} {118}},\
  \bibinfo {pages} {107206} (\bibinfo {year} {2017})}\BibitemShut {NoStop}%
\bibitem [{\citenamefont {Gardner}\ \emph {et~al.}(2010)\citenamefont
  {Gardner}, \citenamefont {Gingras},\ and\ \citenamefont
  {Greedan}}]{Gardner2010}%
  \BibitemOpen
  \bibfield  {author} {\bibinfo {author} {\bibfnamefont {J.~S.}\ \bibnamefont
  {Gardner}}, \bibinfo {author} {\bibfnamefont {M.~J.~P.}\ \bibnamefont
  {Gingras}},\ and\ \bibinfo {author} {\bibfnamefont {J.~E.}\ \bibnamefont
  {Greedan}},\ }\href {https://doi.org/10.1103/RevModPhys.82.53} {\bibfield
  {journal} {\bibinfo  {journal} {Rev. Mod. Phys.}\ }\textbf {\bibinfo {volume}
  {82}},\ \bibinfo {pages} {53} (\bibinfo {year} {2010})}\BibitemShut {NoStop}%
\bibitem [{\citenamefont {den Hertog}\ and\ \citenamefont
  {Gingras}(2000)}]{Den2000}%
  \BibitemOpen
  \bibfield  {author} {\bibinfo {author} {\bibfnamefont {B.~C.}\ \bibnamefont
  {den Hertog}}\ and\ \bibinfo {author} {\bibfnamefont {M.~J.~P.}\ \bibnamefont
  {Gingras}},\ }\href {https://doi.org/10.1103/PhysRevLett.84.3430} {\bibfield
  {journal} {\bibinfo  {journal} {Phys. Rev. Lett.}\ }\textbf {\bibinfo
  {volume} {84}},\ \bibinfo {pages} {3430} (\bibinfo {year}
  {2000})}\BibitemShut {NoStop}%
\bibitem [{\citenamefont {Melko}\ \emph {et~al.}(2001)\citenamefont {Melko},
  \citenamefont {den Hertog},\ and\ \citenamefont {Gingras}}]{Melko2001}%
  \BibitemOpen
  \bibfield  {author} {\bibinfo {author} {\bibfnamefont {R.~G.}\ \bibnamefont
  {Melko}}, \bibinfo {author} {\bibfnamefont {B.~C.}\ \bibnamefont {den
  Hertog}},\ and\ \bibinfo {author} {\bibfnamefont {M.~J.~P.}\ \bibnamefont
  {Gingras}},\ }\href {https://doi.org/10.1103/PhysRevLett.87.067203}
  {\bibfield  {journal} {\bibinfo  {journal} {Phys. Rev. Lett.}\ }\textbf
  {\bibinfo {volume} {87}},\ \bibinfo {pages} {067203} (\bibinfo {year}
  {2001})}\BibitemShut {NoStop}%
\bibitem [{\citenamefont {Bramwell}\ \emph {et~al.}(2001)\citenamefont
  {Bramwell}, \citenamefont {Harris}, \citenamefont {den Hertog}, \citenamefont
  {Gingras}, \citenamefont {Gardner}, \citenamefont {McMorrow}, \citenamefont
  {Wildes}, \citenamefont {Cornelius}, \citenamefont {Champion}, \citenamefont
  {Melko},\ and\ \citenamefont {Fennell}}]{Bramwell2001}%
  \BibitemOpen
  \bibfield  {author} {\bibinfo {author} {\bibfnamefont {S.~T.}\ \bibnamefont
  {Bramwell}}, \bibinfo {author} {\bibfnamefont {M.~J.}\ \bibnamefont
  {Harris}}, \bibinfo {author} {\bibfnamefont {B.~C.}\ \bibnamefont {den
  Hertog}}, \bibinfo {author} {\bibfnamefont {M.~J.~P.}\ \bibnamefont
  {Gingras}}, \bibinfo {author} {\bibfnamefont {J.~S.}\ \bibnamefont
  {Gardner}}, \bibinfo {author} {\bibfnamefont {D.~F.}\ \bibnamefont
  {McMorrow}}, \bibinfo {author} {\bibfnamefont {A.~R.}\ \bibnamefont
  {Wildes}}, \bibinfo {author} {\bibfnamefont {A.~L.}\ \bibnamefont
  {Cornelius}}, \bibinfo {author} {\bibfnamefont {J.~D.~M.}\ \bibnamefont
  {Champion}}, \bibinfo {author} {\bibfnamefont {R.~G.}\ \bibnamefont
  {Melko}},\ and\ \bibinfo {author} {\bibfnamefont {T.}~\bibnamefont
  {Fennell}},\ }\href {https://doi.org/10.1103/PhysRevLett.87.047205}
  {\bibfield  {journal} {\bibinfo  {journal} {Phys. Rev. Lett.}\ }\textbf
  {\bibinfo {volume} {87}},\ \bibinfo {pages} {047205} (\bibinfo {year}
  {2001})}\BibitemShut {NoStop}%
\bibitem [{\citenamefont {Ross}\ \emph {et~al.}(2011)\citenamefont {Ross},
  \citenamefont {Savary}, \citenamefont {Gaulin},\ and\ \citenamefont
  {Balents}}]{Ross2011}%
  \BibitemOpen
  \bibfield  {author} {\bibinfo {author} {\bibfnamefont {K.~A.}\ \bibnamefont
  {Ross}}, \bibinfo {author} {\bibfnamefont {L.}~\bibnamefont {Savary}},
  \bibinfo {author} {\bibfnamefont {B.~D.}\ \bibnamefont {Gaulin}},\ and\
  \bibinfo {author} {\bibfnamefont {L.}~\bibnamefont {Balents}},\ }\href
  {https://doi.org/10.1103/PhysRevX.1.021002} {\bibfield  {journal} {\bibinfo
  {journal} {Phys. Rev. X}\ }\textbf {\bibinfo {volume} {1}},\ \bibinfo {pages}
  {021002} (\bibinfo {year} {2011})}\BibitemShut {NoStop}%
\bibitem [{\citenamefont {Gingras}\ and\ \citenamefont
  {McClarty}(2014)}]{Gingras2014}%
  \BibitemOpen
  \bibfield  {author} {\bibinfo {author} {\bibfnamefont {M.~J.}\ \bibnamefont
  {Gingras}}\ and\ \bibinfo {author} {\bibfnamefont {P.~A.}\ \bibnamefont
  {McClarty}},\ }\href {https://doi.org/10.1088/0034-4885/77/5/056501}
  {\bibfield  {journal} {\bibinfo  {journal} {Reports on Progress in Physics}\
  }\textbf {\bibinfo {volume} {77}},\ \bibinfo {pages} {056501} (\bibinfo
  {year} {2014})}\BibitemShut {NoStop}%
\bibitem [{\citenamefont {Koohpayeh}\ \emph {et~al.}(2008)\citenamefont
  {Koohpayeh}, \citenamefont {Fort},\ and\ \citenamefont
  {Abell}}]{koohpayeh2008}%
  \BibitemOpen
  \bibfield  {author} {\bibinfo {author} {\bibfnamefont {S.}~\bibnamefont
  {Koohpayeh}}, \bibinfo {author} {\bibfnamefont {D.}~\bibnamefont {Fort}},\
  and\ \bibinfo {author} {\bibfnamefont {J.}~\bibnamefont {Abell}},\ }\href
  {https://doi.org/10.1016/j.pcrysgrow.2008.06.001} {\bibfield  {journal}
  {\bibinfo  {journal} {Progress in Crystal Growth and Characterization of
  Materials}\ }\textbf {\bibinfo {volume} {54}},\ \bibinfo {pages} {121}
  (\bibinfo {year} {2008})}\BibitemShut {NoStop}%
\bibitem [{\citenamefont {Koohpayeh}\ \emph {et~al.}(2014)\citenamefont
  {Koohpayeh}, \citenamefont {Wen}, \citenamefont {Trump}, \citenamefont
  {Broholm},\ and\ \citenamefont {McQueen}}]{koohpayeh2014}%
  \BibitemOpen
  \bibfield  {author} {\bibinfo {author} {\bibfnamefont {S.}~\bibnamefont
  {Koohpayeh}}, \bibinfo {author} {\bibfnamefont {J.-J.}\ \bibnamefont {Wen}},
  \bibinfo {author} {\bibfnamefont {B.}~\bibnamefont {Trump}}, \bibinfo
  {author} {\bibfnamefont {C.}~\bibnamefont {Broholm}},\ and\ \bibinfo {author}
  {\bibfnamefont {T.}~\bibnamefont {McQueen}},\ }\href
  {https://doi.org/10.1016/j.jcrysgro.2014.06.037} {\bibfield  {journal}
  {\bibinfo  {journal} {Journal of Crystal Growth}\ }\textbf {\bibinfo {volume}
  {402}},\ \bibinfo {pages} {291 } (\bibinfo {year} {2014})}\BibitemShut
  {NoStop}%
\bibitem [{\citenamefont {Martin}\ \emph {et~al.}(2017)\citenamefont {Martin},
  \citenamefont {Bonville}, \citenamefont {Lhotel}, \citenamefont {Guitteny},
  \citenamefont {Wildes}, \citenamefont {Decorse}, \citenamefont
  {Ciomaga~Hatnean}, \citenamefont {Balakrishnan}, \citenamefont {Mirebeau},\
  and\ \citenamefont {Petit}}]{Martin2017}%
  \BibitemOpen
  \bibfield  {author} {\bibinfo {author} {\bibfnamefont {N.}~\bibnamefont
  {Martin}}, \bibinfo {author} {\bibfnamefont {P.}~\bibnamefont {Bonville}},
  \bibinfo {author} {\bibfnamefont {E.}~\bibnamefont {Lhotel}}, \bibinfo
  {author} {\bibfnamefont {S.}~\bibnamefont {Guitteny}}, \bibinfo {author}
  {\bibfnamefont {A.}~\bibnamefont {Wildes}}, \bibinfo {author} {\bibfnamefont
  {C.}~\bibnamefont {Decorse}}, \bibinfo {author} {\bibfnamefont
  {M.}~\bibnamefont {Ciomaga~Hatnean}}, \bibinfo {author} {\bibfnamefont
  {G.}~\bibnamefont {Balakrishnan}}, \bibinfo {author} {\bibfnamefont
  {I.}~\bibnamefont {Mirebeau}},\ and\ \bibinfo {author} {\bibfnamefont
  {S.}~\bibnamefont {Petit}},\ }\href
  {https://doi.org/10.1103/PhysRevX.7.041028} {\bibfield  {journal} {\bibinfo
  {journal} {Phys. Rev. X}\ }\textbf {\bibinfo {volume} {7}},\ \bibinfo {pages}
  {041028} (\bibinfo {year} {2017})}\BibitemShut {NoStop}%
\bibitem [{\citenamefont {Tang}\ \emph {et~al.}(2020)\citenamefont {Tang},
  \citenamefont {Sakai}, \citenamefont {Kimura}, \citenamefont {Nakamura},
  \citenamefont {Fu}, \citenamefont {Matsumoto}, \citenamefont {Sakakibara},\
  and\ \citenamefont {Nakatsuji}}]{Tang2020}%
  \BibitemOpen
  \bibfield  {author} {\bibinfo {author} {\bibfnamefont {N.}~\bibnamefont
  {Tang}}, \bibinfo {author} {\bibfnamefont {A.}~\bibnamefont {Sakai}},
  \bibinfo {author} {\bibfnamefont {K.}~\bibnamefont {Kimura}}, \bibinfo
  {author} {\bibfnamefont {S.}~\bibnamefont {Nakamura}}, \bibinfo {author}
  {\bibfnamefont {M.}~\bibnamefont {Fu}}, \bibinfo {author} {\bibfnamefont
  {Y.}~\bibnamefont {Matsumoto}}, \bibinfo {author} {\bibfnamefont
  {T.}~\bibnamefont {Sakakibara}},\ and\ \bibinfo {author} {\bibfnamefont
  {S.}~\bibnamefont {Nakatsuji}},\ }\href
  {https://doi.org/10.7566/JPSCP.30.011090} {\bibfield  {journal} {\bibinfo
  {journal} {Proceedings of the International Conference on Strongly Correlated
  Electron Systems (SCES2019)}\ ,\ \bibinfo {pages} {011090}} (\bibinfo {year}
  {2020})}\BibitemShut {NoStop}%
\bibitem [{\citenamefont {Patri}\ \emph {et~al.}(2020)\citenamefont {Patri},
  \citenamefont {Hosoi}, \citenamefont {Lee},\ and\ \citenamefont
  {Kim}}]{Patri2020}%
  \BibitemOpen
  \bibfield  {author} {\bibinfo {author} {\bibfnamefont {A.~S.}\ \bibnamefont
  {Patri}}, \bibinfo {author} {\bibfnamefont {M.}~\bibnamefont {Hosoi}},
  \bibinfo {author} {\bibfnamefont {S.~B.}\ \bibnamefont {Lee}},\ and\ \bibinfo
  {author} {\bibfnamefont {Y.~B.}\ \bibnamefont {Kim}},\ }\href
  {https://doi.org/10.1103/PhysRevResearch.2.033015} {\bibfield  {journal}
  {\bibinfo  {journal} {Phys. Rev. Research}\ }\textbf {\bibinfo {volume}
  {2}},\ \bibinfo {pages} {033015} (\bibinfo {year} {2020})}\BibitemShut
  {NoStop}%
\bibitem [{\citenamefont {Constable}\ \emph
  {et~al.}(2017{\natexlab{a}})\citenamefont {Constable}, \citenamefont
  {Ballou}, \citenamefont {Robert}, \citenamefont {Decorse}, \citenamefont
  {Brubach}, \citenamefont {Roy}, \citenamefont {Lhotel}, \citenamefont
  {Del-Rey}, \citenamefont {Simonet}, \citenamefont {Petit},\ and\
  \citenamefont {deBrion}}]{Constable2017}%
  \BibitemOpen
  \bibfield  {author} {\bibinfo {author} {\bibfnamefont {E.}~\bibnamefont
  {Constable}}, \bibinfo {author} {\bibfnamefont {R.}~\bibnamefont {Ballou}},
  \bibinfo {author} {\bibfnamefont {J.}~\bibnamefont {Robert}}, \bibinfo
  {author} {\bibfnamefont {C.}~\bibnamefont {Decorse}}, \bibinfo {author}
  {\bibfnamefont {J.-B.}\ \bibnamefont {Brubach}}, \bibinfo {author}
  {\bibfnamefont {P.}~\bibnamefont {Roy}}, \bibinfo {author} {\bibfnamefont
  {E.}~\bibnamefont {Lhotel}}, \bibinfo {author} {\bibfnamefont
  {L.}~\bibnamefont {Del-Rey}}, \bibinfo {author} {\bibfnamefont
  {V.}~\bibnamefont {Simonet}}, \bibinfo {author} {\bibfnamefont
  {S.}~\bibnamefont {Petit}},\ and\ \bibinfo {author} {\bibfnamefont
  {S.}~\bibnamefont {deBrion}},\ }\href
  {https://doi.org/10.1103/PhysRevB.95.020415} {\bibfield  {journal} {\bibinfo
  {journal} {Phys. Rev. B}\ }\textbf {\bibinfo {volume} {95}},\ \bibinfo
  {pages} {020415(R)} (\bibinfo {year} {2017}{\natexlab{a}})}\BibitemShut
  {NoStop}%
\bibitem [{\citenamefont {Zhang}\ \emph {et~al.}(2021)\citenamefont {Zhang},
  \citenamefont {Luo}, \citenamefont {Halloran}, \citenamefont {Gaudet},
  \citenamefont {Man}, \citenamefont {Koohpayeh},\ and\ \citenamefont
  {Armitage}}]{Zhang2020}%
  \BibitemOpen
  \bibfield  {author} {\bibinfo {author} {\bibfnamefont {X.}~\bibnamefont
  {Zhang}}, \bibinfo {author} {\bibfnamefont {Y.}~\bibnamefont {Luo}}, \bibinfo
  {author} {\bibfnamefont {T.}~\bibnamefont {Halloran}}, \bibinfo {author}
  {\bibfnamefont {J.}~\bibnamefont {Gaudet}}, \bibinfo {author} {\bibfnamefont
  {H.}~\bibnamefont {Man}}, \bibinfo {author} {\bibfnamefont {S.~M.}\
  \bibnamefont {Koohpayeh}},\ and\ \bibinfo {author} {\bibfnamefont {N.~P.}\
  \bibnamefont {Armitage}},\ }\href
  {https://doi.org/10.1103/PhysRevB.103.L140403} {\bibfield  {journal}
  {\bibinfo  {journal} {Phys. Rev. B}\ }\textbf {\bibinfo {volume} {103}},\
  \bibinfo {pages} {L140403} (\bibinfo {year} {2021})}\BibitemShut {NoStop}%
\bibitem [{\citenamefont {Rau}\ and\ \citenamefont {Gingras}(2019)}]{Rau2019}%
  \BibitemOpen
  \bibfield  {author} {\bibinfo {author} {\bibfnamefont {J.~G.}\ \bibnamefont
  {Rau}}\ and\ \bibinfo {author} {\bibfnamefont {M.~J.}\ \bibnamefont
  {Gingras}},\ }\href
  {https://doi.org/10.1146/annurev-conmatphys-022317-110520} {\bibfield
  {journal} {\bibinfo  {journal} {Annual Review of Condensed Matter Physics}\
  }\textbf {\bibinfo {volume} {10}},\ \bibinfo {pages} {357} (\bibinfo {year}
  {2019})}\BibitemShut {NoStop}%
\bibitem [{\citenamefont {Bonville}\ \emph {et~al.}(2016)\citenamefont
  {Bonville}, \citenamefont {Guitteny}, \citenamefont {Gukasov}, \citenamefont
  {Mirebeau}, \citenamefont {Petit}, \citenamefont {Decorse}, \citenamefont
  {Hatnean},\ and\ \citenamefont {Balakrishnan}}]{Bonville2016}%
  \BibitemOpen
  \bibfield  {author} {\bibinfo {author} {\bibfnamefont {P.}~\bibnamefont
  {Bonville}}, \bibinfo {author} {\bibfnamefont {S.}~\bibnamefont {Guitteny}},
  \bibinfo {author} {\bibfnamefont {A.}~\bibnamefont {Gukasov}}, \bibinfo
  {author} {\bibfnamefont {I.}~\bibnamefont {Mirebeau}}, \bibinfo {author}
  {\bibfnamefont {S.}~\bibnamefont {Petit}}, \bibinfo {author} {\bibfnamefont
  {C.}~\bibnamefont {Decorse}}, \bibinfo {author} {\bibfnamefont {M.~C.}\
  \bibnamefont {Hatnean}},\ and\ \bibinfo {author} {\bibfnamefont
  {G.}~\bibnamefont {Balakrishnan}},\ }\href
  {https://doi.org/10.1103/PhysRevB.94.134428} {\bibfield  {journal} {\bibinfo
  {journal} {Phys. Rev. B}\ }\textbf {\bibinfo {volume} {94}},\ \bibinfo
  {pages} {134428} (\bibinfo {year} {2016})}\BibitemShut {NoStop}%
\bibitem [{\citenamefont {Xu}\ and\ \citenamefont {all}(2021)}]{Xu2020phonons}%
  \BibitemOpen
  \bibfield  {author} {\bibinfo {author} {\bibfnamefont {Y.}~\bibnamefont
  {Xu}}\ and\ \bibinfo {author} {\bibnamefont {all}},\ }\href@noop {}
  {\bibfield  {journal} {\bibinfo  {journal} {manuscript in preparation}\ }
  (\bibinfo {year} {2021})}\BibitemShut {NoStop}%
\bibitem [{\citenamefont {Princep}\ \emph {et~al.}(2013)\citenamefont
  {Princep}, \citenamefont {Prabhakaran}, \citenamefont {Boothroyd},\ and\
  \citenamefont {Adroja}}]{Princep2013}%
  \BibitemOpen
  \bibfield  {author} {\bibinfo {author} {\bibfnamefont {A.~J.}\ \bibnamefont
  {Princep}}, \bibinfo {author} {\bibfnamefont {D.}~\bibnamefont
  {Prabhakaran}}, \bibinfo {author} {\bibfnamefont {A.~T.}\ \bibnamefont
  {Boothroyd}},\ and\ \bibinfo {author} {\bibfnamefont {D.~T.}\ \bibnamefont
  {Adroja}},\ }\href {https://doi.org/10.1103/PhysRevB.88.104421} {\bibfield
  {journal} {\bibinfo  {journal} {Phys. Rev. B}\ }\textbf {\bibinfo {volume}
  {88}},\ \bibinfo {pages} {104421} (\bibinfo {year} {2013})}\BibitemShut
  {NoStop}%
\bibitem [{\citenamefont {Gaudet}\ \emph {et~al.}(2018)\citenamefont {Gaudet},
  \citenamefont {Hallas}, \citenamefont {Buhariwalla}, \citenamefont {Sala},
  \citenamefont {Stone}, \citenamefont {Tachibana}, \citenamefont {Baroudi},
  \citenamefont {Cava},\ and\ \citenamefont {Gaulin}}]{Gaudet2018}%
  \BibitemOpen
  \bibfield  {author} {\bibinfo {author} {\bibfnamefont {J.}~\bibnamefont
  {Gaudet}}, \bibinfo {author} {\bibfnamefont {A.~M.}\ \bibnamefont {Hallas}},
  \bibinfo {author} {\bibfnamefont {C.~R.~C.}\ \bibnamefont {Buhariwalla}},
  \bibinfo {author} {\bibfnamefont {G.}~\bibnamefont {Sala}}, \bibinfo {author}
  {\bibfnamefont {M.~B.}\ \bibnamefont {Stone}}, \bibinfo {author}
  {\bibfnamefont {M.}~\bibnamefont {Tachibana}}, \bibinfo {author}
  {\bibfnamefont {K.}~\bibnamefont {Baroudi}}, \bibinfo {author} {\bibfnamefont
  {R.~J.}\ \bibnamefont {Cava}},\ and\ \bibinfo {author} {\bibfnamefont
  {B.~D.}\ \bibnamefont {Gaulin}},\ }\href
  {https://doi.org/10.1103/PhysRevB.98.014419} {\bibfield  {journal} {\bibinfo
  {journal} {Phys. Rev. B}\ }\textbf {\bibinfo {volume} {98}},\ \bibinfo
  {pages} {014419} (\bibinfo {year} {2018})}\BibitemShut {NoStop}%
\bibitem [{\citenamefont {Sanjurjo}\ \emph {et~al.}(1994)\citenamefont
  {Sanjurjo}, \citenamefont {Rettori}, \citenamefont {Oseroff},\ and\
  \citenamefont {Fisk}}]{Sanjurjo1994}%
  \BibitemOpen
  \bibfield  {author} {\bibinfo {author} {\bibfnamefont {J.~A.}\ \bibnamefont
  {Sanjurjo}}, \bibinfo {author} {\bibfnamefont {C.}~\bibnamefont {Rettori}},
  \bibinfo {author} {\bibfnamefont {S.}~\bibnamefont {Oseroff}},\ and\ \bibinfo
  {author} {\bibfnamefont {Z.}~\bibnamefont {Fisk}},\ }\href
  {https://doi.org/10.1103/PhysRevB.49.4391} {\bibfield  {journal} {\bibinfo
  {journal} {Phys. Rev. B}\ }\textbf {\bibinfo {volume} {49}},\ \bibinfo
  {pages} {4391} (\bibinfo {year} {1994})}\BibitemShut {NoStop}%
\bibitem [{\citenamefont {Thalmeier}\ and\ \citenamefont
  {Fulde}(1982)}]{Thalmeier1982}%
  \BibitemOpen
  \bibfield  {author} {\bibinfo {author} {\bibfnamefont {P.}~\bibnamefont
  {Thalmeier}}\ and\ \bibinfo {author} {\bibfnamefont {P.}~\bibnamefont
  {Fulde}},\ }\href {https://doi.org/10.1103/PhysRevLett.49.1588} {\bibfield
  {journal} {\bibinfo  {journal} {Phys. Rev. Lett.}\ }\textbf {\bibinfo
  {volume} {49}},\ \bibinfo {pages} {1588} (\bibinfo {year}
  {1982})}\BibitemShut {NoStop}%
\bibitem [{\citenamefont {Lovesey}\ and\ \citenamefont
  {Staub}(2000)}]{Lovesey2000}%
  \BibitemOpen
  \bibfield  {author} {\bibinfo {author} {\bibfnamefont {S.~W.}\ \bibnamefont
  {Lovesey}}\ and\ \bibinfo {author} {\bibfnamefont {U.}~\bibnamefont
  {Staub}},\ }\href {https://doi.org/10.1103/PhysRevB.61.9130} {\bibfield
  {journal} {\bibinfo  {journal} {Phys. Rev. B}\ }\textbf {\bibinfo {volume}
  {61}},\ \bibinfo {pages} {9130} (\bibinfo {year} {2000})}\BibitemShut
  {NoStop}%
\bibitem [{\citenamefont {Ruminy}\ \emph {et~al.}(2017)\citenamefont {Ruminy},
  \citenamefont {Chi}, \citenamefont {Calder},\ and\ \citenamefont
  {Fennell}}]{Ruminy2017}%
  \BibitemOpen
  \bibfield  {author} {\bibinfo {author} {\bibfnamefont {M.}~\bibnamefont
  {Ruminy}}, \bibinfo {author} {\bibfnamefont {S.}~\bibnamefont {Chi}},
  \bibinfo {author} {\bibfnamefont {S.}~\bibnamefont {Calder}},\ and\ \bibinfo
  {author} {\bibfnamefont {T.}~\bibnamefont {Fennell}},\ }\href
  {https://doi.org/10.1103/PhysRevB.95.060414} {\bibfield  {journal} {\bibinfo
  {journal} {Phys. Rev. B}\ }\textbf {\bibinfo {volume} {95}},\ \bibinfo
  {pages} {060414(R)} (\bibinfo {year} {2017})}\BibitemShut {NoStop}%
\bibitem [{\citenamefont {Thalmeier}(1984)}]{Thalmeier1984}%
  \BibitemOpen
  \bibfield  {author} {\bibinfo {author} {\bibfnamefont {P.}~\bibnamefont
  {Thalmeier}},\ }\href {https://doi.org/10.1063/1.333518} {\bibfield
  {journal} {\bibinfo  {journal} {Journal of Applied Physics}\ }\textbf
  {\bibinfo {volume} {55}},\ \bibinfo {pages} {1916} (\bibinfo {year}
  {1984})}\BibitemShut {NoStop}%
\bibitem [{\citenamefont {Constable}\ \emph
  {et~al.}(2017{\natexlab{b}})\citenamefont {Constable}, \citenamefont
  {Ballou}, \citenamefont {Robert}, \citenamefont {Decorse}, \citenamefont
  {Brubach}, \citenamefont {Roy}, \citenamefont {Lhotel}, \citenamefont
  {Del-Rey}, \citenamefont {Simonet}, \citenamefont {Petit},\ and\
  \citenamefont {deBrion}}]{Constable2018}%
  \BibitemOpen
  \bibfield  {author} {\bibinfo {author} {\bibfnamefont {E.}~\bibnamefont
  {Constable}}, \bibinfo {author} {\bibfnamefont {R.}~\bibnamefont {Ballou}},
  \bibinfo {author} {\bibfnamefont {J.}~\bibnamefont {Robert}}, \bibinfo
  {author} {\bibfnamefont {C.}~\bibnamefont {Decorse}}, \bibinfo {author}
  {\bibfnamefont {J.-B.}\ \bibnamefont {Brubach}}, \bibinfo {author}
  {\bibfnamefont {P.}~\bibnamefont {Roy}}, \bibinfo {author} {\bibfnamefont
  {E.}~\bibnamefont {Lhotel}}, \bibinfo {author} {\bibfnamefont
  {L.}~\bibnamefont {Del-Rey}}, \bibinfo {author} {\bibfnamefont
  {V.}~\bibnamefont {Simonet}}, \bibinfo {author} {\bibfnamefont
  {S.}~\bibnamefont {Petit}},\ and\ \bibinfo {author} {\bibfnamefont
  {S.}~\bibnamefont {deBrion}},\ }\href
  {https://doi.org/10.1103/PhysRevB.95.020415} {\bibfield  {journal} {\bibinfo
  {journal} {Phys. Rev. B}\ }\textbf {\bibinfo {volume} {95}},\ \bibinfo
  {pages} {020415} (\bibinfo {year} {2017}{\natexlab{b}})}\BibitemShut
  {NoStop}%
\bibitem [{Note1()}]{Note1}%
  \BibitemOpen
  \bibinfo {note} {For $A_{1g}$ singlet level only vibronic interaction with a
  $\Gamma $-point phonon would produce a doublet}\BibitemShut {NoStop}%
\bibitem [{\citenamefont {Trump}\ \emph {et~al.}(2018)\citenamefont {Trump},
  \citenamefont {Koohpayeh}, \citenamefont {Livi}, \citenamefont {Wen},
  \citenamefont {Arpino}, \citenamefont {Ramasse}, \citenamefont {Brydson},
  \citenamefont {Feygenson}, \citenamefont {Takeda}, \citenamefont {Takigawa}
  \emph {et~al.}}]{Trump2018}%
  \BibitemOpen
  \bibfield  {author} {\bibinfo {author} {\bibfnamefont {B.~A.}\ \bibnamefont
  {Trump}}, \bibinfo {author} {\bibfnamefont {S.~M.}\ \bibnamefont
  {Koohpayeh}}, \bibinfo {author} {\bibfnamefont {K.~J.}\ \bibnamefont {Livi}},
  \bibinfo {author} {\bibfnamefont {J.-J.}\ \bibnamefont {Wen}}, \bibinfo
  {author} {\bibfnamefont {K.}~\bibnamefont {Arpino}}, \bibinfo {author}
  {\bibfnamefont {Q.~M.}\ \bibnamefont {Ramasse}}, \bibinfo {author}
  {\bibfnamefont {R.}~\bibnamefont {Brydson}}, \bibinfo {author} {\bibfnamefont
  {M.}~\bibnamefont {Feygenson}}, \bibinfo {author} {\bibfnamefont
  {H.}~\bibnamefont {Takeda}}, \bibinfo {author} {\bibfnamefont
  {M.}~\bibnamefont {Takigawa}}, \emph {et~al.},\ }\href
  {https://doi.org/10.1038/s41467-018-05033-7} {\bibfield  {journal} {\bibinfo
  {journal} {Nature communications}\ }\textbf {\bibinfo {volume} {9}},\
  \bibinfo {pages} {2619} (\bibinfo {year} {2018})}\BibitemShut {NoStop}%
\end{thebibliography}%

\end{document}